# Understanding the Origin of the Low Cure-Shrinkage of Polybenzoxazine Resin by Computational Simulation

Prashik S. Gaikwad[1], Aaron S. Krieg[1], Prathamesh P. Deshpande[1], Sagar Umesh Patil[1], Julia A. King[1], Marianna Maiaru[2], Gregory M. Odegard[1]

[1]Michigan Technological University, Houghton, 49931, USA
[2]University of Massachusetts Lowell, Lowell, 01854, USA

## ABSTRACT

Thermoset resin-based composite materials are widely used in the aerospace industry, mainly due to their high stiffness-to-weight and strength-to-weight ratios. A major issue with the use of thermoset resins in fiber composites is the process-induced residual stresses that are formed from resin chemical shrinkage during the curing process. These residual stresses within the composite material ultimately result in reduced durability and residual deformations of the final product. Polybenzoxazine (PBZ) polymer resins have demonstrated near-zero volumetric shrinkage during the curing process. Although the low shrinkage of PBZ is promising in terms of reduced process-induced residual stresses, little is known about the physical causes. In this work, Molecular Dynamics (MD) simulations are performed with a reactive force field to predict the physical properties (gelation point, evolution of network, mass density, volumetric shrinkage) and mechanical properties (Bulk modulus, Shear modulus, Young's Modulus, Poisson's ratio, Yield strength) as a function of crosslinking density. The MD modeling procedure is validated herein using experimental measurements of the modeled PBZ resin. The results of this study are used to provide a physical understanding of the zero-shrinkage phenomenon of PBZ. This information is also a critical input to future process modeling efforts for PBZ composites.

## 1. INTRODUCTION

Polymer matrix composites (PMCs) are widely used in the aerospace industry because of their high strength/weight and high stiffness/weight ratios [1]. Thermoset-based PMCs undergo chemical crosslinking when subjected to elevated temperatures and pressures, thus converting the resin from a liquid to a solid phase. During this transformation, the polymer undergoes chemical and thermal shrinkage. During the processing of thermoset PMCs, this shrinkage leads to the development of residual stresses within the material, especially near the reinforcement interface. Furthermore, residual stresses also result from the thermal contraction mismatch between the resin matrix and reinforcement during the cool-down cycle of PMC processing. These combined residual stresses can lead to mis-shaped final products; as well as the formation of matrix microcracks and premature fiber/matrix debonding, thus reducing the overall product durability [2]. The evolution of these residual stresses as a functional of processing parameters can be computationally predicted using process modeling [3].

Polybenzoxazine (PBZ) resins are high-performance thermosets that have demonstrated excellent thermal, mechanical, chemical, electrical, and physical properties. In particular, they have a high char yield, high glass transition temperature ($T_g$), low water absorption, low flammability, and near-zero volume shrinkage during curing [4]. The latter property is of particular interest in producing composite laminates with minimal residual stresses. Process modeling can be used to predict the influence of the evolving thermal, mechanical, and physical properties (including near-zero shrinkage) on the accumulation of residual stresses. However, accurate process modeling requires the resin properties as a function of crosslinking. Thus, a comprehensive understanding of the evolution of the properties of a particular resin is required to optimize the processing parameters of the composite. Experimentally determining the evolution of properties of a resin as a function of crosslinking can be very time-consuming and expensive. Fortunately, Molecular Dynamics (MD) simulation can be used to accurately and efficiently predict the mechanical properties of thermosets during curing [5].

Several computational studies have been carried out to study various benzoxazine resin systems. In a molecular simulation by Kim et al. [6], a fully relaxed atomistic molecular model of amorphous polybenzoxazine was developed using Polymer Consistent force field (PCFF) [7] to determine the cohesive energy density, atomic radial distribution function, distribution of torsional angles, and correlation between phenyl rings and the free volume fraction as a function of radius of probe. Kim et al. [8] further studied the surface properties of polybenzoxazines thin films. Thompson et al. [9] used MD to estimate the glass transition temperature ($T_g$) of five benzoxazine polymer systems. An MD study was carried out by Saiev et al. [10] and used the Dreiding force field [11] to gain fundamental insight into the relationship between structural network topology and the thermomechanical properties of three PBZ systems. A new crosslinking protocol with simulation was developed by Sanders et al. [12] simulated the crosslinking of benzoxazine and predicted mass density and volumetric shrinkage using the OPLS2005 force field [13, 14, 15]. Despite these efforts to computationally simulate PBZ systems, the evolution of structure and mechanical properties as a function of crosslink density is yet to be comprehensively explored, and a clear physical explanation for their near-zero cure shrinkage is yet to be provided. Furthermore, the MD simulation of PBZ systems with a reactive force field for accurate physical and mechanical property predictions has not yet been documented.

The aims of this study are to accurately predict the physical and mechanical properties of PBZ resin as a function of crosslinking density at room temperature using MD simulation and to provide physical insight into the zero-shrinkage nature of PBZ. In this study, the Reactive Interface Force Field (IFF-R) is used to accurately predict the physical and mechanical properties for large deformations. The predictions are validated through experiments. The results provide a comprehensive set of properties for process modeling and a physical description of the curing process.

## 2. MATERIAL SYSTEM

For this study, the Bisphenol – A benzoxazine (Ba-BZ) system was modeled. Figure 1a shows the molecular structure of the Ba-BZ monomer (commercially marketed as Araldite 35600).

## 3. MOLECULAR MODELING

The MD simulations were performed using the LAMMPS (Large-scale Atomic/Molecular Massively Parallel Simulator) MD software package. The Reactive Interface Force Field (IFF-R) [16] was used to describe interatomic forces, as implemented in LAMMPS [17, 18], using the parametrization described in Odegard et al. [19]. IFF-R is a reactive force field which is a modified version of the Polymer Consistent Force Field – Interface Force Field (PCFF-IFF) [20,21], with the harmonic covalent bond stretch potentials replaced with Morse potentials. The use of Morse potentials allows for the more accurate calculation of bond stretch energies at higher deformations, including the possibility of bond dissociation. Recently, Odegard et al. [19] implemented IFF-R for predicting thermo-mechanical properties of epoxies which closely matched experimental and literature values, and Patil et al. [22] used IFF-R to predict the evolution of physical and mechanical properties of epoxy during curing.

## 3.1 Model densification and equilibration

A single Ba-BZ monomer was replicated in the x,y,z directions of a MD simulation box, which consisted of a total of 14,040 atoms (256 monomers). The model was slowly compressed (densified) to a target mass density of 1.20 g/cc using the NPT (fixed pressure and temperature) ensemble. The densification simulation was run for 5 nanoseconds (ns) at room temperature (300 K) and 1 atm with 1 femtoseconds (fs) time steps. After densification, an annealing simulation was performed to allow the molecules to achieve more desirable configurations. In this step, the temperature was ramped from 300 K to 600 K and then slowly cooled from 600 K to 300 K at a 50K/ns cooling rate with the NVT (fixed volume and temperature) ensemble. The annealed model was further subjected to an equilibration simulation at 300K and 1 atm for 2 ns with 1 fs timesteps. The purpose of the relaxation step was to allow the model to relax further and prepare for crosslinking. These steps were repeated five times to establish five independent replicates to account for statistical deviations in the predicted properties. The Nose-Hoover thermostat and barostat were implemented for all the simulations discussed herein. Figure 1b,1c shows the minimized Ba-BZ structure and equilibrated MD model of PBZ respectively.

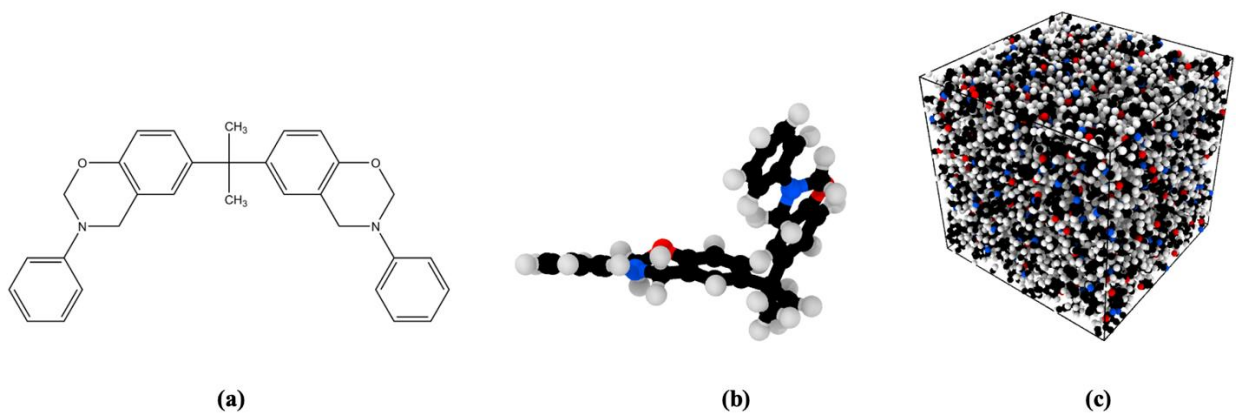

**Figure 1:** (a) Molecular structure of Ba-BZ, (b) Minimized structure of Ba-BZ, (c) Equilibrated MD model of PBZ.

## 3.2 Crosslinking of Benzoxazine Monomers

A single-crystal X-ray crystallographic study shows that the preferential conformation of the oxazine ring is a distorted semi-chair structure [23]. Polymerization of Ba-BZ takes place via a thermally-induced ring-opening reaction, thus resulting in a phenolic structure characterized by a Mannich base bridge ($-CH_2-N-CH_2-$) [24]. Chutayothin et al. [25] performed a detailed analysis of the final polymerization products, and by combining experimental results with the basicity of amine substituent in the benzoxazine monomer, the initiation mechanism of the benzoxazine system was determined. Liu et al. [26] proposed an improved mechanistic scheme for ring-opening polymerization of benzoxazines using Hydrogen Nuclear Magnetic Resonance analysis. Rucigaj et al. [27] studied the influence of phenolic, mercapto, amino, and imidazole accelerators on the curing of benzoxazine.

In this study, thermally-activated ring opening crosslinking without accelerators was simulated. Figure 2 shows the crosslinking of Ba-BZ. Crosslinking of Ba-BZ is a two-step process. In the first step, the covalent bond between the aliphatic carbon and the oxygen in the oxazine ring breaks (Figure 2a), thus opening the oxazine ring as shown in Figure 2b. In the second step, a covalent bond is formed between the aliphatic carbon of the oxazine ring and the ortho- aromatic carbon of the other Ba-BZ molecules, resulting in a Mannich base bridge ($-CH_2-N-CH_2-$) as shown in Figure 2c. Similar, ring opening crosslinking occurs at the other side of Ba-BZ as shown in Figure 2d. Figure 2e shows the finalized crosslinked structure of PBZ.

The LAMMPS command "fix bond/react" developed by Gissinger et al. [28] was used to simulate the crosslinking steps. Seven different crosslinked models were created ranging from 10% to 70% with increments of 10%. The crosslinking density is defined as the ratio of total number of new covalent bonds that are formed between the monomers to the maximum total number of covalent bonds that can be formed. These crosslinking simulations were performed at 650 K with 0.1 fs timesteps using the NVT ensemble. The crosslinking simulations for each crosslinked model ran for 500 ps. Sequential crosslinking was used [22] for building the full range of crosslinked models. The crosslinking procedure was used for each of the five replicates.

After crosslinking, each model was annealed at 650 K for 7 ns with 1 fs timesteps using the NPT ensemble to further drive the molecular structures to more desirable configurations. During annealing, the temperature was ramped down from 650 K to 300 K at a 50K/ns cooling rate. The models were then equilibrated at 300 K and 1 atm for 2 ns with 1 fs timesteps using the NPT ensemble. During these equilibration simulations, the volume and density of the simulation boxes were tracked. This information was used to predict the mass density and the volumetric shrinkage of each crosslinked model. For all these simulations, the Nose-Hoover barostat was set to maintain a pressure of 0.101 MPa (1 atm) on all sides of the simulation box.

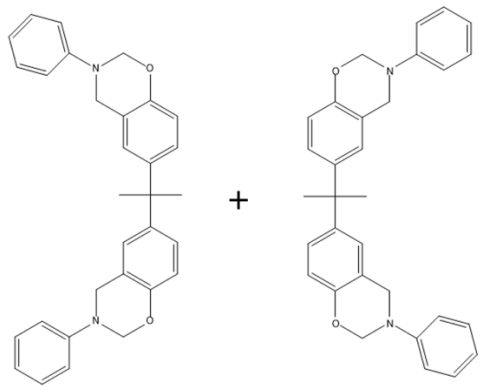

(a)

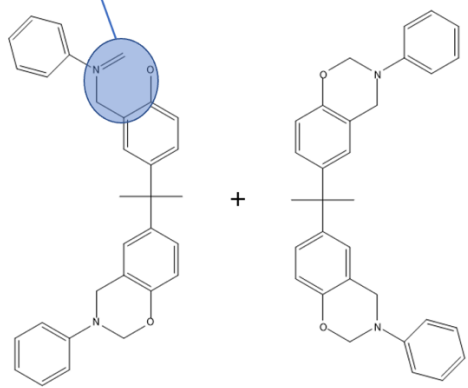

(b)

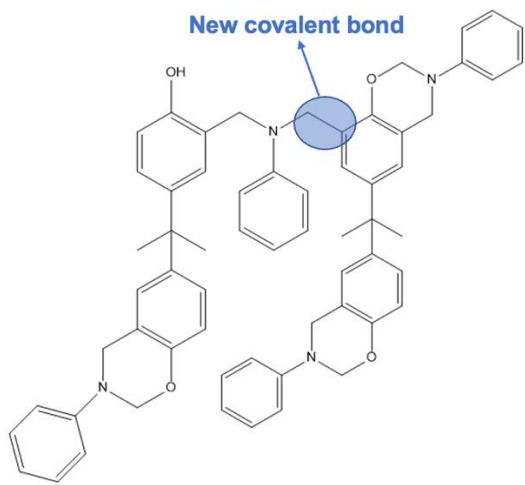

(c)

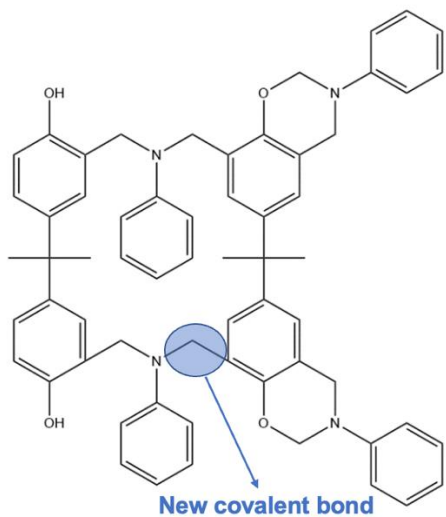

(d)

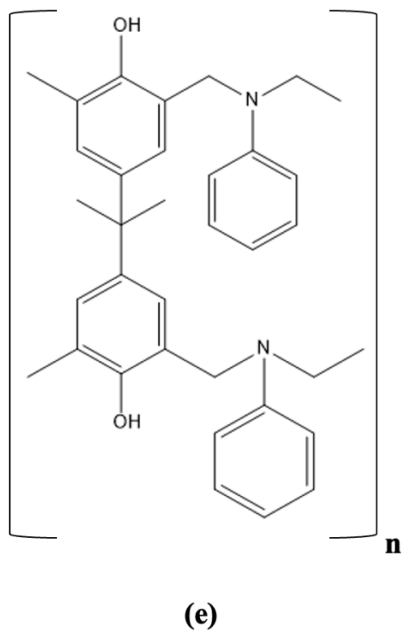

**Figure 2:** Crosslinking of Ba-BZ. (a) Two individual BA-BZ monomers initiate the ring-opening polymerization. (b) Ring-opening polymerization occurs and the covalent bond between the O atom and the C atom in the oxazine ring breaks. (c) Due to the oxazine ring opening, a new covalent bond forms, thus leading to the Mannich bridge structure (-CH$_2$-N-CH$_2$-). (d) Same as (c) step, but at a second location. (e) Final polymerized structure, where n represents the number of polymeric units.

## 3.3 Gel point

Polymer gelation occurs when an infinite crosslinked networked forms, that is, when the entire system becomes a single molecule [29]. Even though the thermoset polymer is not fully crosslinked at the gel point, the covalent bond network can bear significant mechanical loads applied in any direction. The polymerization gel point can be predicted using MD by tracking the evolution of molecular weights of the independent polymer clusters during the crosslinking process [5, 30, 31]. Figure 3 shows the predicted molecular weight of the PBZ clusters as a function of crosslinking density. An in-house Python script was used to calculate the molecular weights of the PBZ clusters. Three different metrics can be used to quantify the gel point. First, the figure shows the evolution of the molecular weight of the largest cluster, and the corresponding inflection point (shown in Figure 3) can be used to quantify the gel point, which is at a crosslinking density of 45%. Second, the peak of the molecular weight of the second-largest cluster is another means of identifying the gel point, which is at 40%. Finally, the peak of the weight-averaged molecular weight (RMW) of all the reacting groups except for the largest group is another method of determining the gel point, with a corresponding value of 40%. From the results of the three metrics, the gel point of the PBZ resin is in the range of 40% - 45% crosslinking density. 40% crosslinking density was chosen as the gel point of this resin system for calculating post-gelation volumetric shrinkage (described below).

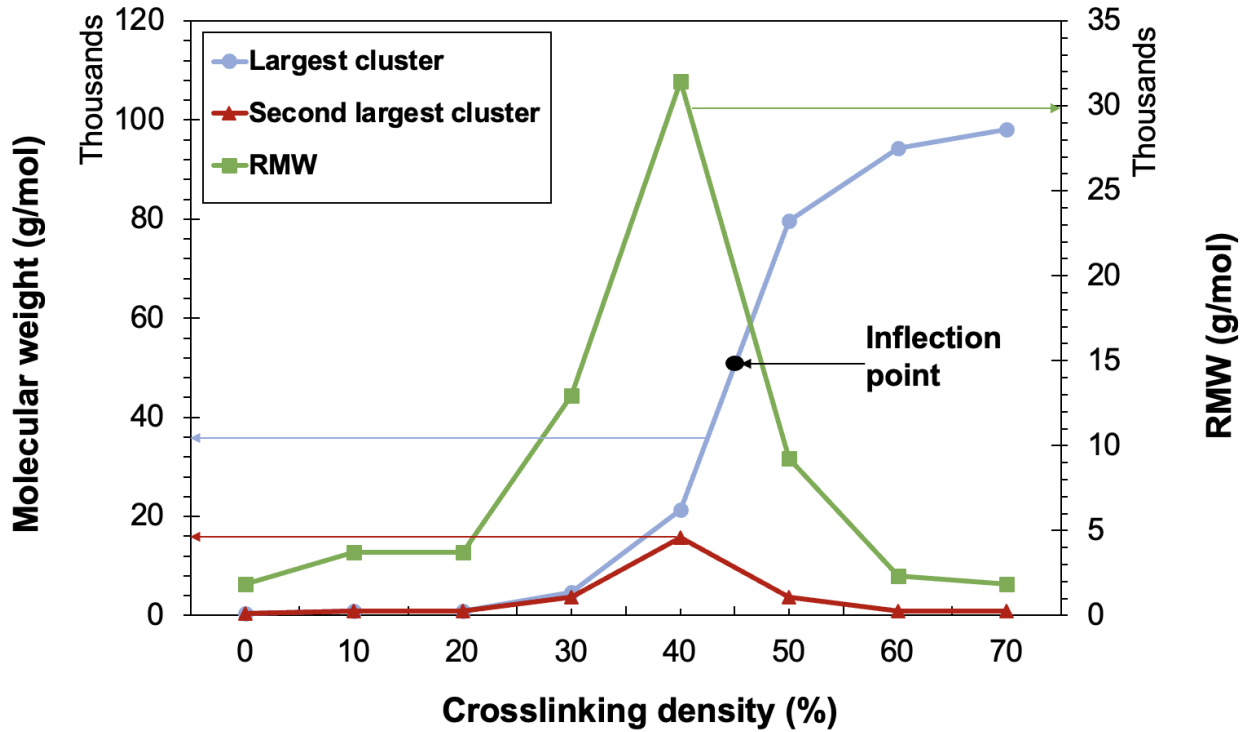

**Figure 3:** Molecular weight evolution of PBZ as a function of crosslinking density.

## 3.4 Evolution of network topology

At any point in time during the curing process, the monomer/polymer mixture consists of three types of clusters: free chains, dangling chains, and crosslinked chains. Free chains are monomers which are not connected to any polymer chains. Dangling chains consist of dimers, trimers, and oligomers that are not connected to the largest chain in the system. The crosslinked chain has the highest molecular weight fraction in the system. This chain is the main contributor to the elastic behavior of the material. Free and dangling chains do not significantly contribute to the elastic behavior [5].

Figure 4 shows the chain weight percent of the free, dangling, and crosslinked chain as a function of crosslinking density for a representative MD replicate model. The number of free chains decreases as the crosslinking density increases from 0% to 70%. At low crosslinking densities (0% to 30%), the number of dangling chains increases and reaches a peak value at 30%, and progressively decreases thereafter. As the crosslinking proceeds further, the number of crosslinked chains increases by consuming the dangling and free chains. At the maximum crosslinked density (70%), the system has formed a complete network and nearly every polymer unit can participate in load transfer along the network of covalent bonds.

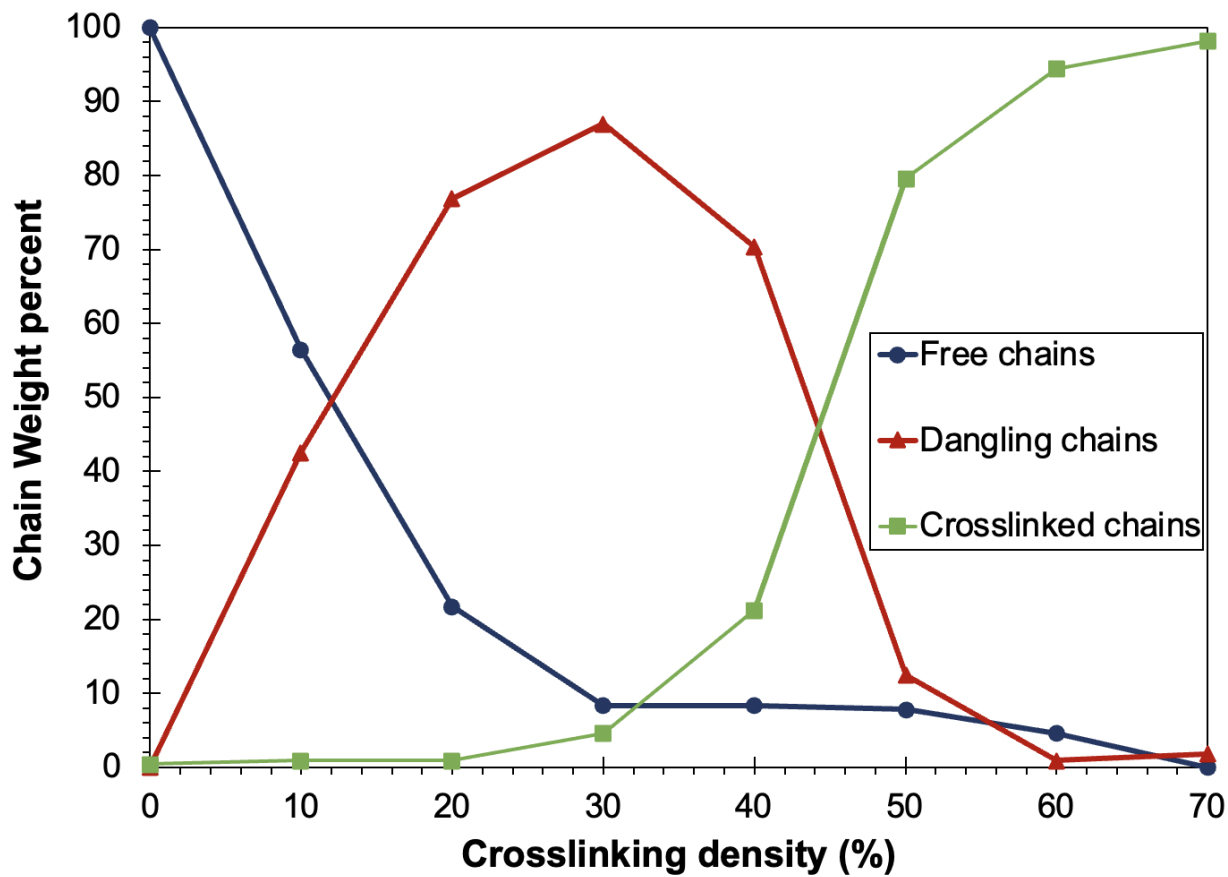

**Figure 4:** Evolution of clusters as a function of crosslinking density for a representative MD model.

Figure 5 shows the evolution of the polymer network in a representative MD simulation box at crosslinking densities of 0%, 20%, 40%, 60%, and 70%. The Open Visualization Tool (OVITO) software package [32] was used for obtaining the images shown in Figure 5. Each color in the images represents an individual cluster. The 0% crosslinked model consists of all free chains. As the crosslink density increases from 0 to 20%, the number of dangling chains increases. At a crosslink density of 40% (the gel point), there is a clear formation of the crosslinked chain (represented by the color green). Above 40% crosslinking density, the size of the crosslinked chain increases dramatically, while the number of free and dangling chains decreases. At a 70% crosslinked density, the crosslinked chain completely spans the box.

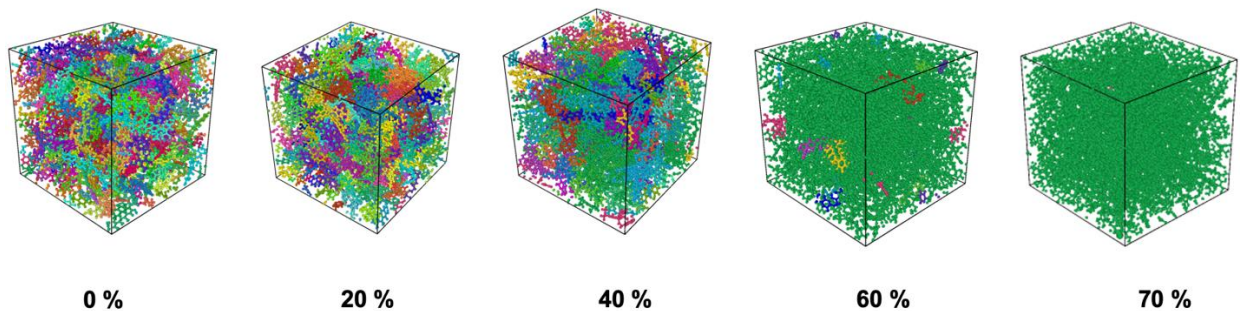

**Figure 5:** Snapshots of clusters at varying crosslinking densities.

## 3.6 Physical properties

The volumetric shrinkage was calculated for each replicate and crosslink density using:

$$\Delta V = \frac{V_f - V_o}{V_o} \qquad (1)$$

where $V_f$ is the volume of the simulation box at a specific crosslink density (10% to 70%) and $V_o$ is volume of the simulation box for the fully uncrosslinked (0%) system. The post-gelation volumetric shrinkage is calculated using

$$\Delta V_{pg} = \frac{V_{gf} - V_g}{V_g} \qquad (2)$$

where $V_{gf}$ is volume of the simulation box at a specific crosslink density (> 40%) in the post gelation region and $V_g$ is the volume of the simulation box at the gel point (40%).

## 3.7 Mechanical properties

For bulk modulus predictions, MD simulations were run at room temperature (300 K), 1 atm for 1 ns at 1 fs timesteps using the NPT ensemble. The pressure was subsequently increased to 5000 atm and the simulations were run for an additional 1 ns at 300 K using the NPT ensemble. By computing the change in the equilibrated volume for each of the two pressures, the bulk modulus was calculated for each crosslinked density and replicate using

$$K = -V_{eq}\left(\frac{dP}{dV}\right) \qquad (3)$$

Where $V_{eq}$ is the average volume of the simulated box recorded during equilibrium process. To predict the shear modulus, shear deformations were individually applied to the simulation boxes in all three principal planes. A shear strain of 20% was applied at a strain rate of $2 \times 10^8$ s$^{-1}$. These simulations were carried out at room temperature (300 K) and at 1 atm pressure using the NPT ensemble with 1 fs time steps. For the shear stress-strain curve associated with each replicate, crosslink density, and principal plane, a bilinear breakpoint was determined by observing the strain at which the slope changes dramatically. The slope of the line below the breakpoint of the shear stress-shear strain graph was calculated as the shear modulus. Further details of the shear modulus calculations are described by Odegard et al. [19].

Assuming the material to be isotropic, the Young's modulus (*E*) and Poisson's ratio (*v*) for each of the models were calculated from the corresponding values the *K* and *G* using the standard isotropic elasticity equations [19]:

$$E = \frac{9KG}{3K + G} \qquad (4)$$

$$n = \frac{3K - 2G}{2(3K + G)} \quad (5)$$

The uniaxial yield strength was also determined using the data from the shear simulations. The von Mises stress was calculated from the individual stress components while shearing the box individually in all three principal planes using

$$S_{vM} = \sqrt{\frac{1}{2}\left[(S_x - S_y)^2 + (S_y - S_z)^2 + (S_z - S_x)^2 + 6(t_{xy}^2 + t_{xz}^2 + t_{yz}^2)\right]} \quad (6)$$

The corresponding yield strength was the von Mises stress value at the bilinear breakpoint. Further details of the yield strength calculation are discussed by Odegard et al. [19]. The yield strength was determined for each replicate, shearing plane, and crosslink density.

## 4. Experimental study

This section details the fabrication and testing of specimens for validation of the MD results. Each method is described in an individual sub-section.

### 4.1. Neat PBZ plate fabrication

Araldite 35600 PBZ samples were manufactured using a compression molding method. Two speedmixer cups were charged with 70 g each of the resin. Speedmixer cups were heated to 120°C and mixed in a FlackTek Speedmixer DAC 150.1 FVZ at 2500 rpm for 3 minutes. The Ba-BZ resin was cast into a tooling assembly and compression molded at 180 °C for 2 hours and then ramped to 200 °C and held for 2 hours. The compression molder was cooled using air and water until the system was cooled to 150 °C and then was switched to water cooling only to continue to cool the system to 25°C before removing the plate. The tooling assembly produced two 152 mm×152 mm plates with 3.2 mm thickness.

### 4.2. Mass density measurement

PBZ mass density samples were prepared by melting the resin to 110 °C and casting into puck shaped molds. The Ba-BZ resin was then degassed for 45 minutes at 110 °C. Samples were then removed from the oven and allowed to cool overnight to 23 °C. The mass density of the cured resin was determined according to ASTM D792, which uses the buoyancy force of a sample of known mass submerged in a liquid of known density to calculate the mass density of the sample. Test samples were cut from a plate using a vertical band saw so that the geometries of test samples were 38.1 mm long, 12.7 mm wide, and 3.2 mm thick.

### 4.3. Mechanical testing

Specimens were tested for tensile properties at 23°C according to ASTM D638 using the ASTM Type I sample geometry: 165.1 mm long and 3.2 mm thick. A Ceast router was used to grind the specimens into dog-bone shaped samples with a width of 12.6 mm. Eight samples were tested at a crosshead rate of 1 mm/min using an Instron 4206 screw-driven mechanical testing machine. Stress values were recorded by the testing machine and a 50.8 mm axial extensometer from Epsilon Technology Corporation was used to collect elongation data from which the axial strain was calculated. The tensile modulus was determined from the initial slope of the stress-strain curve. The samples were conditioned at 23 °C and 50% relative humidity for two days prior to testing.

## 5. Results and Discussion

Figure 6 shows the mass density as a function of crosslinking density at room temperature. As the crosslinking density increases from 0 % to 70 %, the mass density values remain nearly constant, as indicated by the uncertainty bars in Figure 6. These results clearly show that during the curing process of PBZ, there is little change in the mass density, which agrees with reports of zero cure shrinkage for this resin system [33]. From Figure 6 it is also evident that that the MD predicted mass density of 1.182 ± 0.001 g/cc of the maximum crosslinked system (70 %) agrees very well with the experimental density value of 1.182 ± 0.002 g/cc (n=10) measured in this study along with the literature mass density values [33, 34].

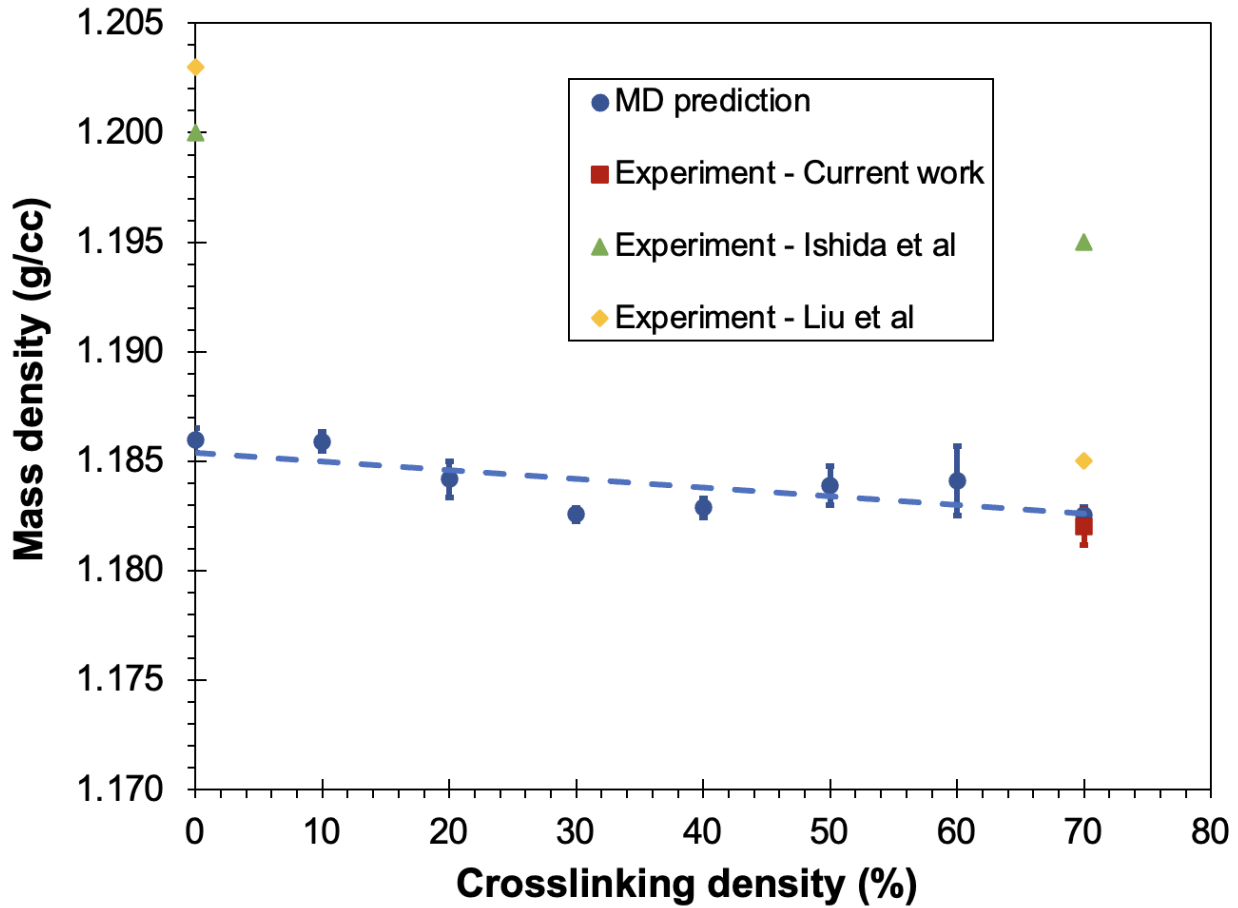

**Figure 6:** Mass density as a function of crosslinking density. MD prediction values are averages of five replicates and the error bars are the corresponding standard error. The MD predicted mass density value for the highest crosslinked system is compared with the experimental measured values from Ishida et al. [33] and Liu et al. [34]. The data points are fitted using a linear function to guide the eye.

Figure 7 shows the volumetric shrinkage as a function of crosslinking density. It is clear that there is a small negative volumetric shrinkage as the crosslinking density increases from 0% to 70%, which is opposed to the typical shrinkage behavior of epoxies and other thermosets, which typically show a significant positive shrinkage [22]. The inset in Figure 7 shows the volumetric shrinkage in the post-gelation regime. In this region, a small positive shrinkage (contraction of resin) is observed. However, the range of the positive shrinkage is less than 0.5 % which is insignificant relative to other thermoset resins [22].

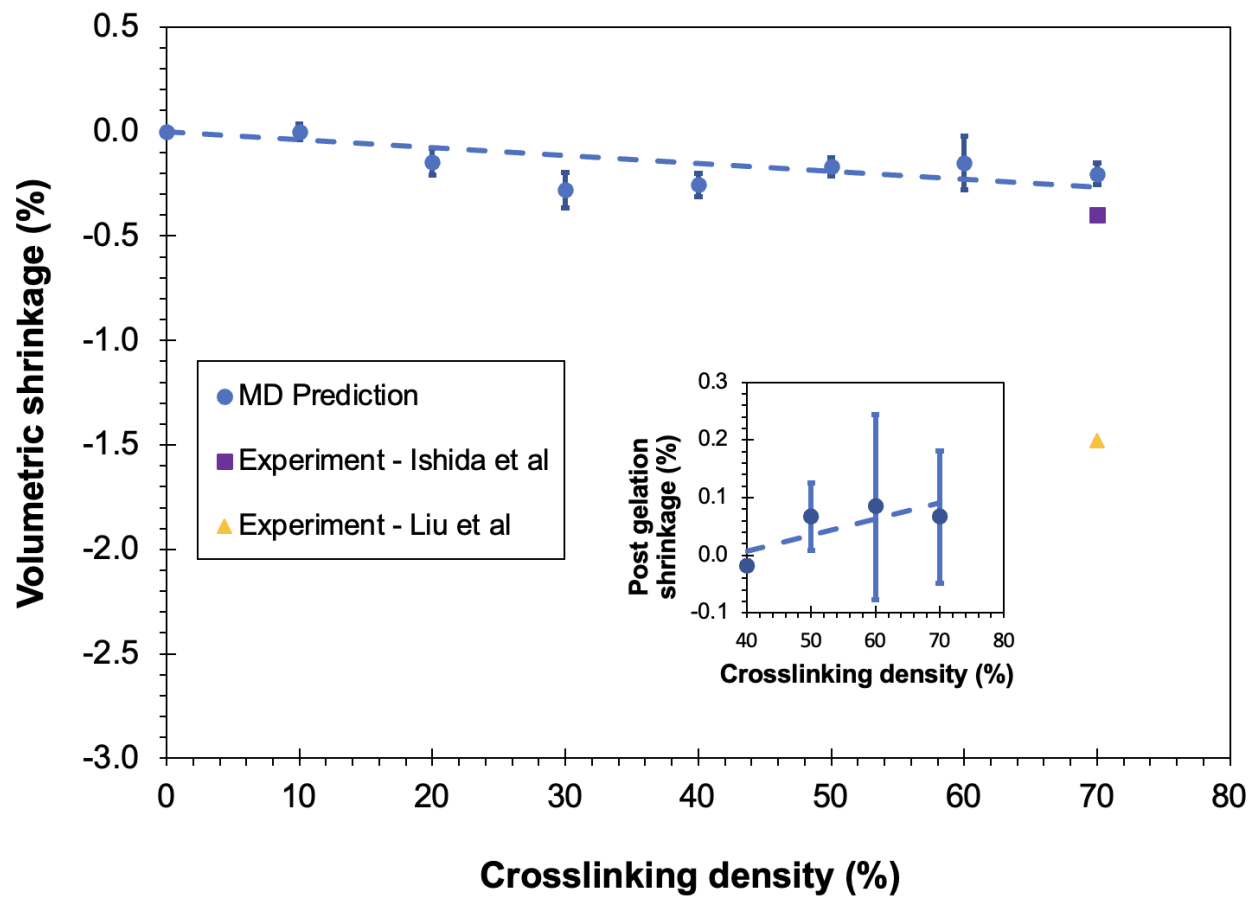

**Figure 7:** Volumetric shrinkage as a function of crosslinking density. Post-gelation shrinkage as a function of crosslinking density (inset). Each data point represents the average of five MD models and the vertical error bars represent standard error over the replicates. The data points are fitted using a linear function to guide the eye.

Figure 8 shows the radial distribution function of phenyl rings for the uncrosslinked (0% crosslinking density) and max crosslinked (70% crosslinking density) models of PBZ. Included in the figure is also the radial distribution functions for two other systems: bisphenol-A epoxy with an amine hardener [35] and bisphenol-F epoxy with an amine hardener [36] .The two epoxy systems have been included because they exhibit significant shrinkage during crosslinking [35, 36] and thus provide physical insight into the near-zero shrinkage behavior of PBZ. From the figure it is clear that the alignment/spacing of phenyl groups in the PBZ system changes very little during crosslinking. Thus, it follows that the overall molecular structure changes very little during the cure process. Specifically, the liquid resin consists of monomers with close spacings that do not appreciably change as the crosslinks are formed. This is in contrast with the two epoxy systems, which show significant change in the phenyl group spacing during cure as the epoxy monomers are pulled closer together as they chemically bond with amines. It was previously speculated [37] that the lack of cure shrinkage in PBZ was due to the formation of hydrogen bonds during the crosslinking process. However, it is unclear how the formation of such bonds contributes to a state of zero shrinkage, especially since the uncrosslinked state already has a relatively high mass density when compared to other structural thermoset resins, which is likely due to the tight packing of PBZ monomers in the uncured state.

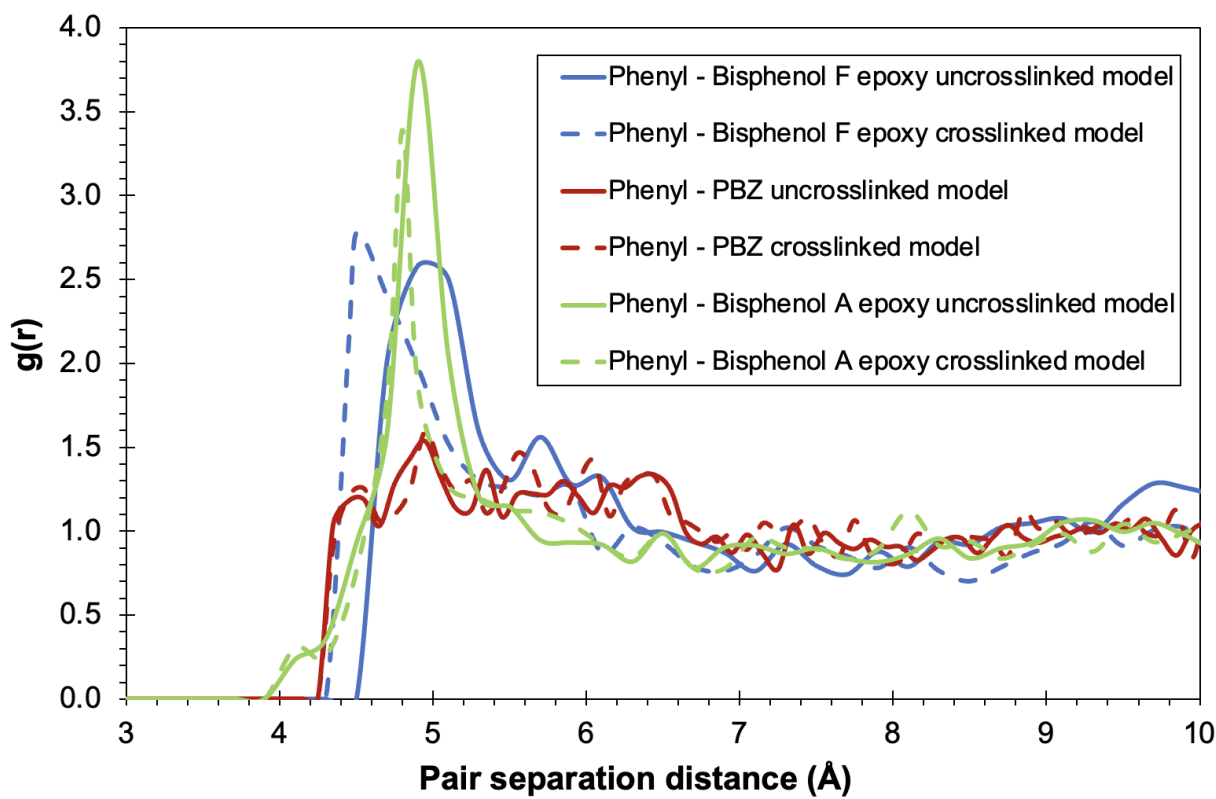

**Figure 8:** Radial distribution function of phenyl rings for the uncrosslinked and the maximum crosslinked models.

The evolution of shear modulus and bulk modulus as a function of crosslinking density is shown in Figure 9. A bulk modulus of $6.02 \pm 0.03$ GPa and shear modulus of $2.03 \pm 0.01$ GPa was predicted for the max crosslinked system (70%). An increase in the shear and bulk modulus is observed with increasing crosslinking density. This is expected because the formation of crosslinks serves to create a more extensive covalent bond network that can transfer significant mechanical loads. There does not appear to be a significant change in the trends of shear and bulk moduli before and after the gel point.

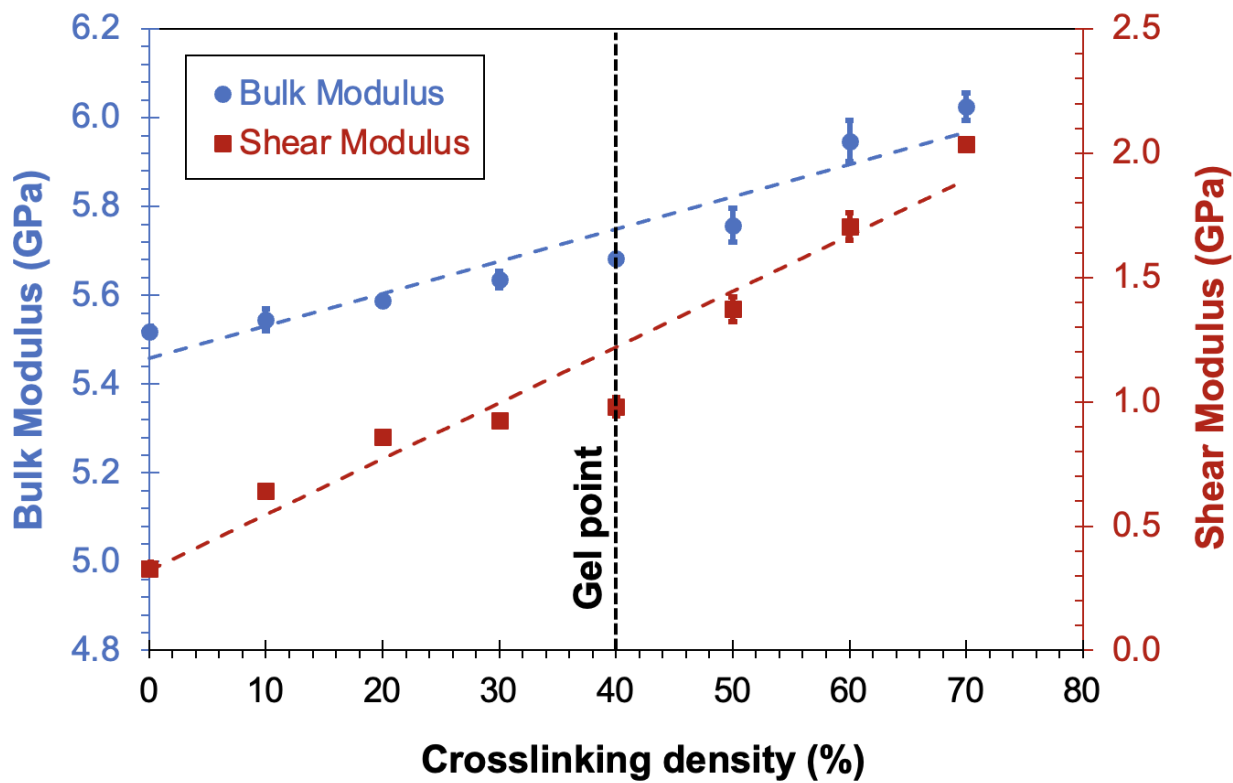

**Figure 9:** Shear modulus (red) and bulk modulus (blue) as a function of crosslinking density. Each data point represents the average of five MD models and the vertical error bars represent the standard error over the replicates. The data points are fitted using a linear function to guide the eye.

Figure 10 shows the evolution of the Young's modulus and Poisson's ratio as a function crosslinking. A steady increase in the Young's Modulus is observed as the crosslinking density increases. The MD predicted Young's modulus value at the fully crosslinked state of 5.58 ± 0.10 GPa agrees well with the experimental measurement of 5.49 ± 0.33 GPa. Although a significant discrepancy between Young's modulus predictions from MD simulation and experimental measurements is typically observed for epoxies [19, 38, 39, 40, 41], the same discrepancy is not observed here with the PBZ system. As discussed by Odegard et al. [19] for epoxies, this strain rate effect is caused by a significant viscous mechanical response in the epoxies. However, this particular PBZ is much stiffer and has a significantly higher glass transition temperature than typical epoxies. Thus, it is expected to have a relatively small viscous response in terms of modulus, which explains the lack of an observable strain rate effect in Figure 10.

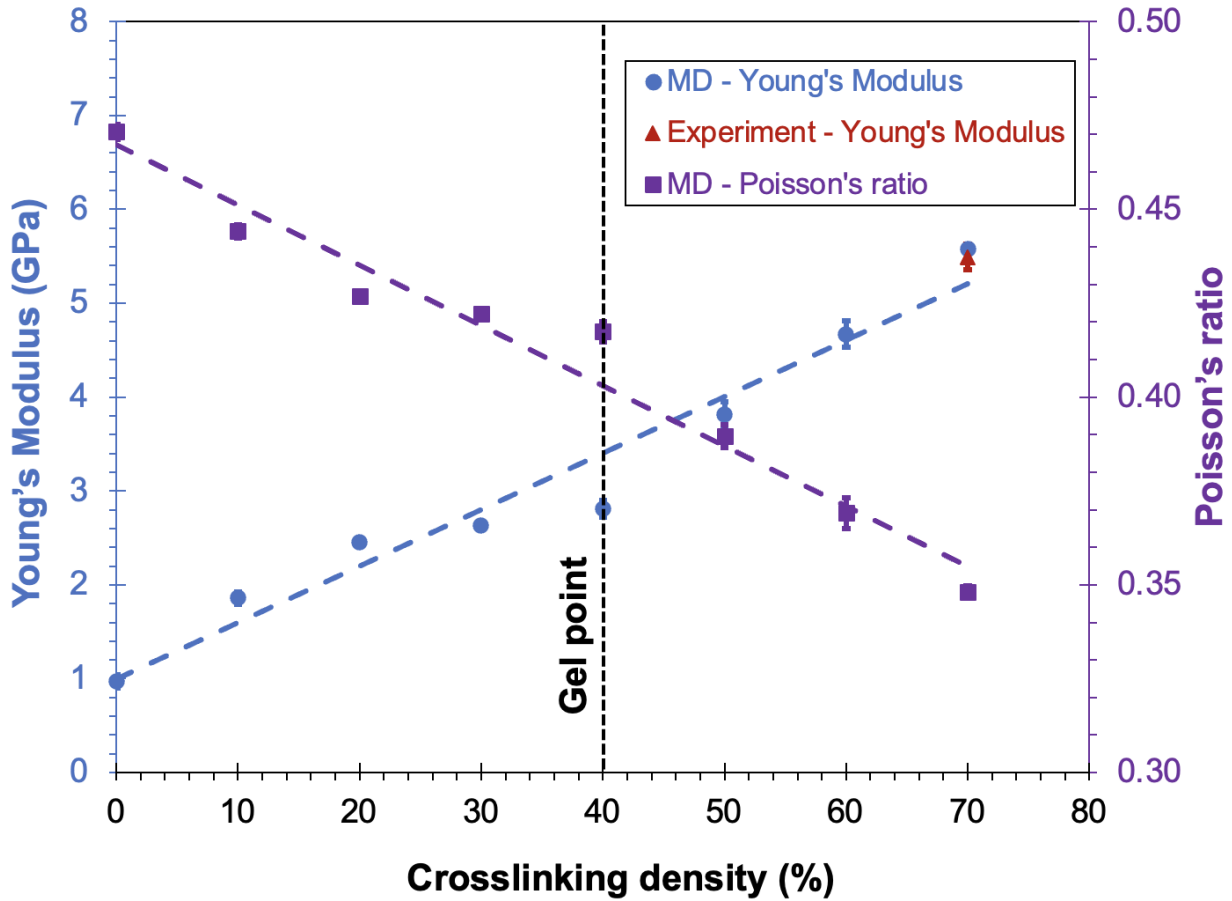

**Figure 10:** Young's modulus and Poisson's ratio as a function of crosslinking density. Each data point represents the average of five MD models and the vertical error bars represent standard error over the replicates. The MD predicted data points are fitted using a linear function to guide the eye.

Figure 10 shows a decrease in the Poisson's ratio as the crosslinking density increases, which is likely due to the formation of a stronger crosslinked network during curing that reduces the transverse response of the material to axial deformations. A Poisson's ratio of 0.348 ± 0.001 was predicted for the max crosslinked system (70 %).

Figure 11 shows the yield strength as a function of crosslinking density. Overall, an increase in the yield strength is observed as the crosslinking density increases, which is expected as the formation of a more highly networked system should correspond to an increasing resistance to yielding. A yield strength of 131.98 ± 3.95 MPa was predicted using MD for the max crosslinking density (70%). In Figure 11 the MD predicted value for the max crosslinked density is compared to the experimental measured value of 49.2 ± 2.02 MPa (at a strain rate of $3.25 \times 10^{-4}$ s$^{-1}$). The significant discrepancy between the MD predicted value and experimentally measured value is due to the strain-rate effect, which has a more significant impact on yield strength predictions than modulus predictions [19, 38, 42].

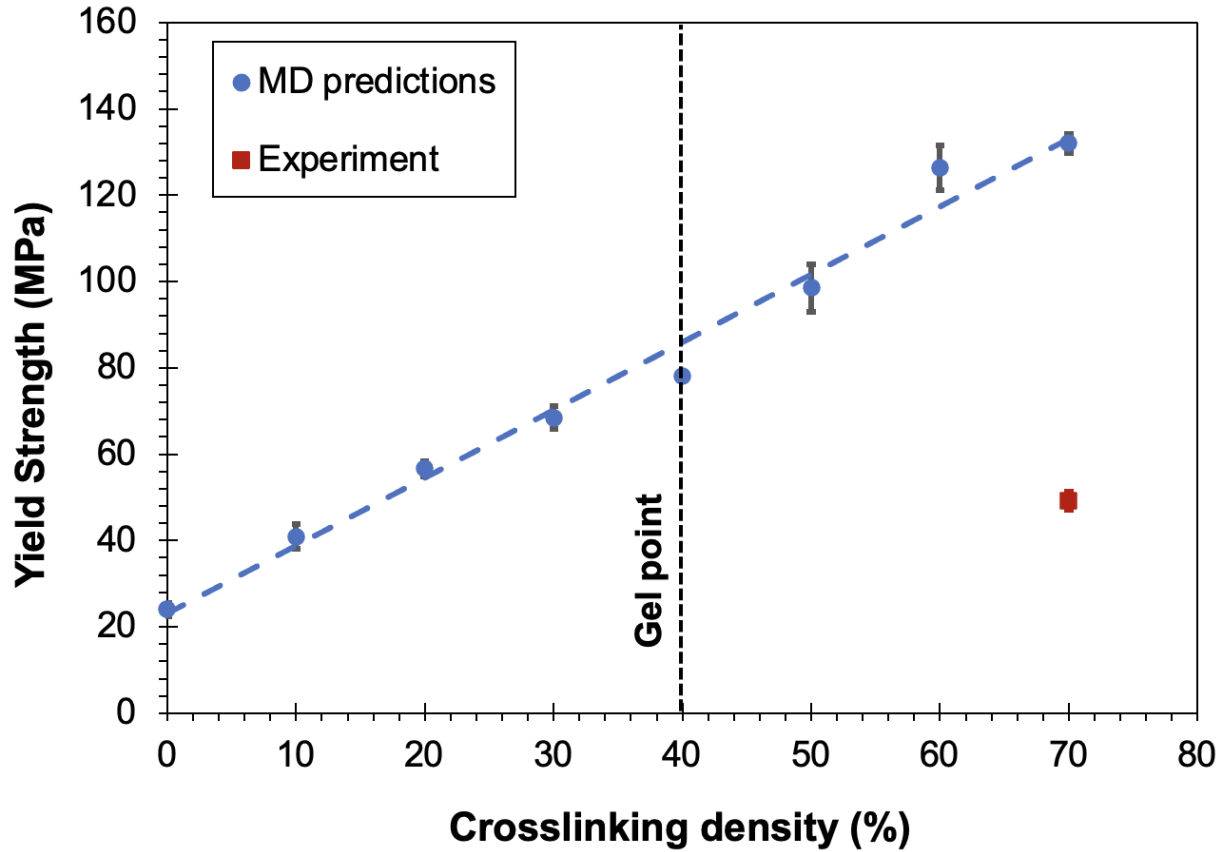

**Figure 11:** Yield strength as a function of crosslinking density. Each MD data point represents the average of five MD models and the vertical error bars represent the standard error over the replicates. The MD predicted data points are fitted using a linear function to guide the eye.

## 5 Conclusions

There are several important points to make regarding the results of this study. First, the prediction of physical and mechanical properties generally shows good agreement with experimental measurements, which indicates that the modeling procedure and IFF-R both provide accurate predicted properties for the PBZ system. This is further evidence that IFF-R is an excellent force field to use for the prediction of physical and mechanical properties of thermosetting polymers [19, 22].

Second, the results provide a deeper understanding into the zero-shrinkage nature of the PBZ resin. Because the uncrosslinked system with efficiently packed monomers undergoes very little configurational change during the crosslinking process, a negligible change in the overall mass density is observed during the network formation process. Thus, the use of pure PBZ resin in composites processing can mitigate some of the internal residual stresses that typically result from chemical shrinkage of the curing thermoset.

Finally, the predicted properties provide a quantitative understanding of their evolution during the curing process. This information is critical for optimizing composites processing parameters, reducing post-manufacturing residual stresses, and improving the quality and strength of composite laminate structures.

## Acknowledgements

This research was partially supported by the NASA Space Technology Research Institute (STRI) for Ultra-Strong Composites by Computational Design (US-COMP), grant NNX17AJ32G; and NASA grant 80NSSC19K1246. SUPERIOR, a high-performance computing cluster at Michigan Technological University, was used in obtaining the MD simulation results presented in this publication.